# THE SPIRAL1 CHARGE BREEDER: KEY POINTS FOR A PERFORMANT 1+ BEAM INJECTION

L. Maunoury[†1], O. Kamalou[1], J.F. Cam[2], S. Damoy[1], P. Delahaye[1], M. Dubois[1], R. Frigot[1], S. Hormigos[1] and C. Vandamme[2]

[1]Grand Accélérateur National d'Ions Lourds, Bd Henri Becquerel, 14076 Caen, France

[2]Laboratoire Physique Corpusculaire, 6 Bd Maréchal Juin, 14050 Caen, France

*Abstract*

The SPIRAL1 charge breeder is now under operation. Radioactive beam has already been delivered [1] to Physicist for performing experiment. Although charge breeding efficiencies demonstrated high performances for stable ion beams, those efficiencies regarding radioactive ion beams were found, in the first experiments, lower than expected. The beam optics, prior to the injection of the 1+ ions into the SPIRAL1 charge breeder, is of prime importance [2] for getting such high efficiencies. Moreover, the intensities of the radioactive ion beams are so low, that it is really difficult to tune the charge breeder. The tuning of the charge breeder for radioactive ion beams requires a particular procedure often referred as "blind tuning". A stable beam having a close Brho (few %) is required to find out the set of optic parameters preceding the tuning of the radioactive beam. Hence, it has been decided to focus our effort on that procedure as to get control of the 1+ beam optics leading to high charge breeding efficiencies whatever the 1+ mass, energy and Target Ion Source System (TISS) used. Being aware that each TISS provide ion beams with a specific energy spread ΔE, and given that the acceptance energy window of the charge breeder is rather narrow; that parameter must play also an important role in the whole charge breeding efficiency.

This contribution will show the strategy undertaken to overcome the difficulties encountered in the charge breeding tuning with 1+ radioactive ion beams from the different ion sources, and the results already obtained. A series of experiment have been done to record beam parameters as well as beam profiles in two modes: shooting through and 1+/n+. Simulations have been developed to replicate the measurements; SIMION as well as TraceWin are the two software that were combined together for that purpose. The final goal is to define a set of beam optics reliable enough for operation covering a large range of 1+ mass, energy and emittances; applying that set of parameters must eventually leadclose to expected best achievable charge breeder performances while producing a radioactive ion beam.

## INTRODUCTION

Now, the SPIRAL1 facility has been upgraded and is under operation. Obviously, more time is needed to get a full control over the entire facility as it got more complicated to operate: new Target ion Source System (TISS) [3]; a charge breeder (SP1CB) [4] and reshuffled beam lines. Mainly there are two operational modes for running the facility providing Radioactive Ion Beams (RIB's) to physicist: Shooting Through mode (ST) and 1+/N+ mode [5]. The ST mode mainly applies to beams to the NanoganIII TISS [6] providing RIB's of gas element type transported to the post-accelerator CIME cyclotron [7] through the 6 mm plasma electrode of the SP1CB. The 1+/N+ mode combines the FEBIAD TISS with the SP1CB and has successfully provided a $^{38m}$K@9MeV/u beam [8] to physicists in 2019, after first charge breeding results were obtained in 2018 with $^{37}$K [9]. But challenges ought to be overcome for having a daily routine facility. Indeed, the preparing phase prior to the RIB delivery is long and difficult, and a unique set of beam optics parameters must be found for each mode. A dedicated R&D is therefore important to understand what are key parameters of the ion beam (Twiss parameters and energy spread) upstream the SP1CB in order to achieve the highest charge breeding efficiencies.

## THE 1+ BEAM LINE

### 1+ beam line layout

Figure 1 displays the layout of the 1+ beam line. It starts with a combination of two einzel lenses: Einz1 and Einz2. It is followed by a triplet of magnetic quadrupoles and a magnetic sextupole prior to a magnetic mass analyzing dipole. Next, a triplet of electrostatic quadrupoles is settled just before the 1+ ion injection into the SP1CB. For the diagnostics, there are two beam profilers (PR11 and PR13, respectively upstream and downstream the mass analyzing dipole), sets of slits (vertical - horizontal) and one faraday cup CF13, shielded regarding the ion beam extracted backwards from the SP1CB. After the second mass analyzing dipole D3P sorting out the multi-charged ions extracted from the SP1CB, a Faraday cup with a beam profiler is used to check the selected multi charged ion beam intensity and its profile.

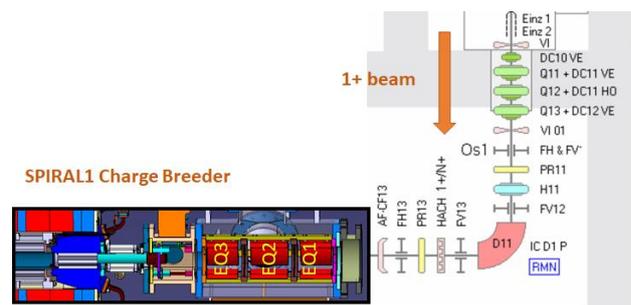

† laurent.maunoury@ganil.fr

Figure 1: Layout of the 1+ beam line.

*Shooting Through mode (ST)*

In this mode, the TISS NanoganIII is operated. The specificity is the large emittance of that Electron Cyclotron Resonance (ECR) ion source type. Moreover, the ion beam must be transported across the SP1CB plasma electrode of 6 mm diameter coercing it to be really focused at that location. The experimental data (beam profilers and efficiency of transport) of a stable ion beam of $^{14}N^{3+}$@19.751kV have been considered. As the twiss parameters for the transverse emittances are not known for the NanoganIII ECR ion source, they were left as free parameters. Using the TraceWin code [10], the ion beam envelopes as well as the initial beam twiss parameters are obtained by adjusting the simulation input to reproduce experimental profiles. The corresponding beam envelope best reproducing the experimental data is shown on Figure 2. The ion beam is focused, thanks to the coils of the charge breeder, and transported as well as possible across the SP1CB plasma electrode. The calculated transverse emittances of the initial beam has been determined to be around 106 π.mm.mrad (4 RMS, X and Y planes), it is a diverging ion beam. But, to replicate the profiler values, a factor of 0.6 on the complete magnetic field intensity of the SP1CB was applied in TraceWin. Another discrepancy was found in the N+ line (the beam line transporting the multi-charged ion beam from the SP1CB plasma electrode up to the CIME injection), where a burst of the ion beam envelope in X plane cannot be really explained. To find out the reason for this discrepancy, a pepper-pot emittance meter (Pantechnik type) will be installed in the N+ beam line to constrain the TraceWin simulation. New data will be recorded and compared to the simulations.

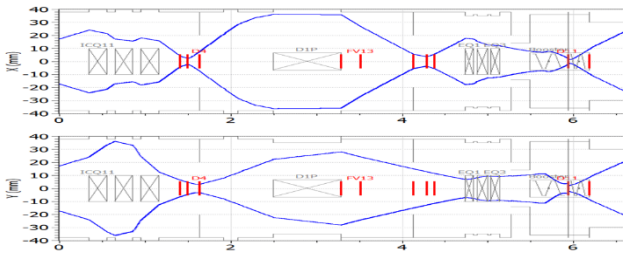

Figure 2: Transverse beam envelopes for the ST mode and a beam $^{14}N^{3+}$@19.751kV.

*Beam optics for 1+/N+ mode*

For many years, the TISS NanoganIII has been used in combination with a unique set of beam optics suitable for elements up to Bρ of 0.136 T.m. Following the upgrade of the SPIRAL1 facility, it is possible nowadays to get RIB's characterized with a Bρ$_{max}$ of 0.22 T.m. The enlargement of Bρ involves some complications regarding beam optics element in the 1+ beam line. Indeed, the magnetic quadrupoles Q11 Q12 and Q13 cannot be simply scaled up from previous beam optics parameters, as the new required field value exceeds their limits. Henceforth, the idea is to find out another set of beam optics parameters matching the injection of the SP1CB to maintain as well as possible an optimized charge breeding efficiency over the whole range of Bρ.

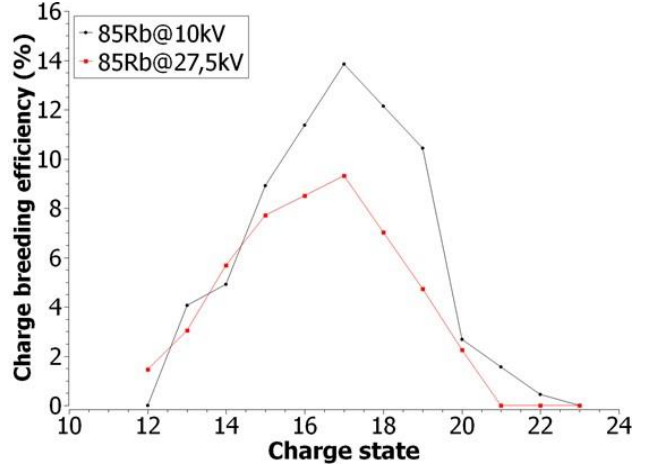

Figure 3: SPIRAL1 charge breeding efficiencies for the mode 1+/N+ and the two cases: $^{85}Rb^+$@10kV (black curve) and $^{85}Rb^+$@27.5kV (red curve).

A $^{85}Rb^{1+}$ ion beam was used for this optimization as at different Bρ's, correspond to different 1+ ion beam energies. We studied the two cases: $^{85}Rb^{1+}$@10kV – Brho = 0.133 T.m (previous beam optics parameters) and $^{85}Rb^{1+}$@27.5kV – Brho = 0.221 T.m. (new beam optics parameters). The $^{85}Rb^{1+}$ ion beam is generated thanks to an ion gun developed at GANIL [2] using a HeatWave pellet [11], delivering up to several hundreds of nA. Figure 3 shows the SP1CB efficiencies for those two cases.

Both charge state distributions are close in FWHM (5 - 6 charge states). They are both peaked on the $^{85}Rb^{17+}$ with charge breeding efficiencies of 13.8% and 9.3% associated with global efficiencies of 70.3% and 49.8% for 10 keV and 27.5 keV 1+ beams, respectively. At the same time, beam profiler values have been recorded to coerce the simulations.

*Simulations and Ion beam injection into the Charge Breeder*

Simulations have been undertaken to find out the new beam optics as well as to learn how the 1+ particles are injected into the SP1CB. For that purpose, two softwares have been operated: SIMION [12] and TraceWin [10] to do a complete simulation of the total SPIRAL1 beam line. Simulations were achieved in a simplified approach with no plasma model implemented in the SP1CB.

SIMION is employed to generate the ion beam from the ion gun up to the exit of the second einzel lens Einz2. One example is displayed on Figure 4. The geometry included in the simulation is as close as possible to the real one and the voltages applied are exactly those used during experiment.

The twiss parameters are deduced and used as input for the TraceWin code. Inside TraceWin, a tool box has to be defined for reproducing each of the different types of optical elements. The beam line has been built to reproduce the 1+ and N+ beam line of the SPIRAL1 facility. For the magnetic as well as for the electrostatic quadrupoles, mass analyzing dipole and einzel lenses, in-house boxes are set in the simulation; gradient (T.m) or voltages are applied to the quadrupoles and the ion beam Bρ is used to define the ion beam trajectory inside the mass analyzing dipoles. Finally, evolution of the transverse beam envelopes (in both planes X and Y) can be calculated as well as the beam profiler values. On Figure 5, the similated beam profiler values in red are compared to the experimental ones in italics and blue. As it can be seen, values are very close except for the X dimension of PR13. In that case, the experimental beam profile was polluted by a huge background.

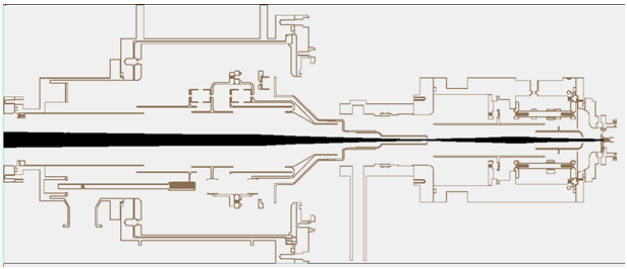

Figure 4: SIMION simulation of 85Rb+@10kV.

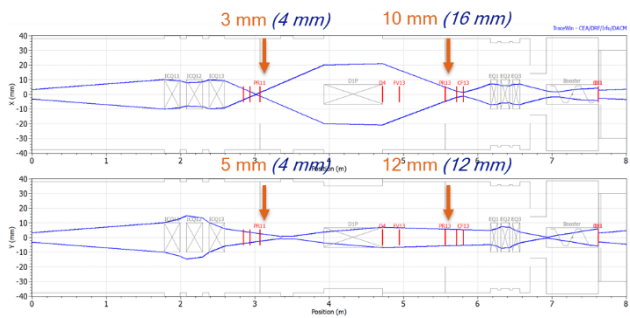

Figure 5: Transverse beam envelopes for the case 1+/N+ mode and the beam $^{85}Rb^+$@10kV.

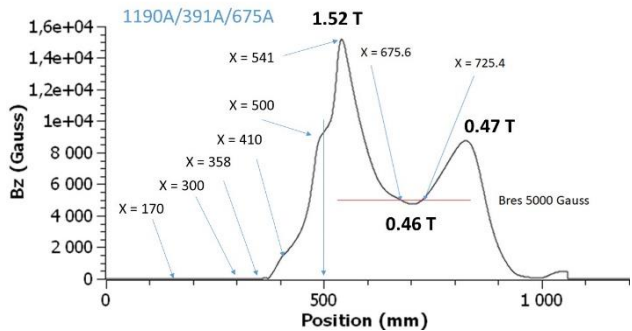

Figure 6: Axial magnetic field profile (Bz in Gauss) with the longitudinal positions corresponding to those shown on Figures 7, 8 and 10.

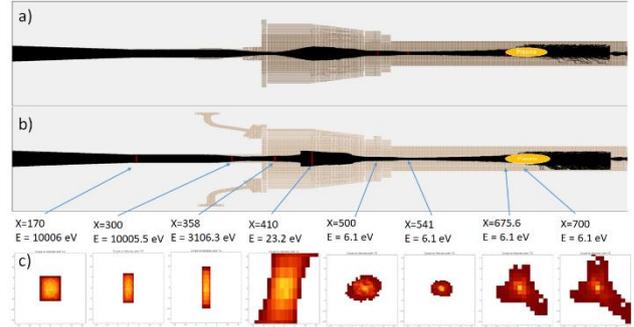

Figure 7: Evolution of the 85Rb+@10kV injected and slowed down into the SP1CB. a) and b) are the pile-up of 1+ trajectories in both transverse planes. c) images representing the evolution of the 2D transverse 1+ beam profile along the SP1CB.

In Figure 5 the ion beam is firstly divergent and is focused in both transverse planes to the object plane of the mass analyzing dipole (D1P). After the mass analyzing dipole, the ion beam is relatively large in vertical plane but again well focused in horizontal plane to get a good resolution. The electrostatic quadrupole focuses in both planes the ion beam to be transported along the deceleration tube of 28 mm of inner diameter. Figure 6 shows the axial magnetic field and the positions where the 2D transverse 1+ beam profile has been calculated. Figure 7 exhibits how the ion beam propagates inside the SP1CB, without plasma. From the TraceWin calculation, twiss parameters are obtained just after the EQ3. These parameters are the inputs to the next SIMION simulation for SP1CB injection, which performs the last step of the simulation: evolution of the 1+ beam injected into the charge breeder and slowing down. The real 3D magnetic field (coils and hexapole), also containing the fringe field, is included into the SIMION simulation  At the entrance of the charge breeder, the ion beam is smoothly convergent to pass through the deceleration tube into the SP1CB. Till the magnetic field is null, ion beam squeezes (X = 170, 300, 358). As soon as the ions feel SP1CB magnetic field, it broadens with a maximum at X = 410 corresponding to the highest magnetic field at injection, pinching thereafter at X = 500, 541 to turn out to be very small. Entering the plasma core zone (X = 675.6, 700), the monocharged particles of the ion beam are magnetically controlled due to their low energy (star shape owing to the combination of magnetic field produced by hexapole and coils). In this region, they should collide with charge particles ($X^{q+}$) of the plasma to be slowed down and captured. The collisions have been not accounted for in this simulation. Figure 8 shows the same evolution depending on the electromagnetic field in use. The "E+B" case is identical to the one of Figure 7 c). The "E" case corresponds to the exclusion of the magnetic field and the "B" case represents the withdrawal of the deceleration electric field. From the cases "E" and "B", it is demonstrated that only the combination of both electromagnetic fields leads to the formation of a tiny ion beam transporting the 1+ ions at the right energy (few eV) up to the core of the SP1CB plasma to be efficiently captured and charge bred.

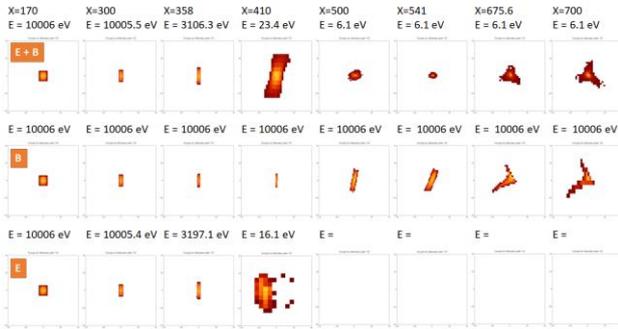

Figure 8: Evolution of the $^{85}$Rb$^+$@10kV injected into the SP1CB for three cases: E+B (both electric field and magnetic field); B (electric field is removed); E (magnetic field is removed).

Same method can be applied to the case of $^{85}$Rb$^+$@27.5kV. But a new beam optics has been found to overcome the magnetic quadrupole (Q11 – Q12 – Q13) limitations. I The $^{85}$Rb$^+$@10kV and $^{85}$Rb$^+$@27.5kV are characterized by Bρ's of 0.133 T.m and 0.221 T.m respectively. A simple scaling with respect to the Bρ's would imply to multiply the currents applied to the magnetic elements by a factor 1.66 (ratio of square root of the acceleration voltages $\sqrt{\frac{27.5}{10}}$). The new currents exceeding the limits of the magnetic quadrupole, a new beam optics has been calculated and tested experimentally. Figure 3 displays the charge state distribution (CSD) of charge bred ions (red curve) at this new energy, with the new beam optics. The new beam optics is shown on Figure 9. It is similar to the one of Figure 5. The most important differences are:
- for the X beam envelope, the global shape is quite comparable except at the entrance of the SP1CB where the beam dimension blows up
- for Y beam envelope, it is generally larger than in Fig. 5 , in particular when passing through the EQ2, before being pinched at the entrance of the SP1CB

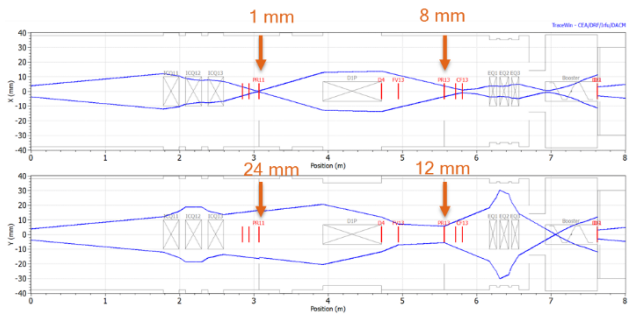

Figure 9: Transverse beam envelopes for the case 1+/N+ mode and the beam $^{85}$Rb$^+$@27.5kV.

Despite these differences, the beam width values on the beam profiler PR13 are quite similar for both cases, prior to the injection inside the SP1CB: for the X plane the widths are 10 mm for 10kV to be compared to 8 mm for 27.5kV and for the Y plane 12 mm for 10kV to be compared to 12 mm for 27.5kV. Figure 10 displays an equivalent behavior compared to Figure 7. The 2D transverse profiles have comparable shapes at the same locations. But, as the beam envelopes in Fig. 9 are larger in the Y plane than those of Fig. 5(cf. discussion above), it looks likewise in Figure 10, for which a larger beam is observed compared to Fig. 7. As the beam in Y plane is a bit larger than the deceleration tube inner diameter, there are some losses which can explain the decrease of the global efficiency in that case (Figure 3 red curve).

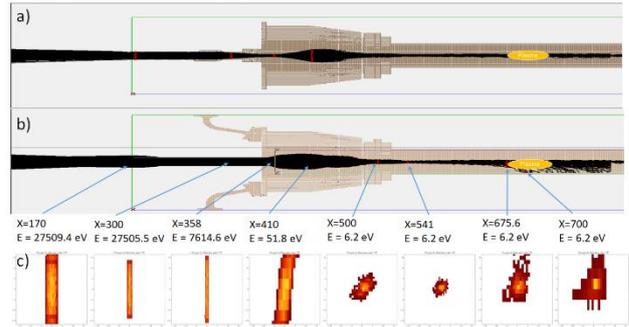

Figure 10: Evolution of the $^{85}$Rb$^+$@27.5kV injected and slowed down into the SP1CB. a) and b) are the pile-up of 1+ trajectories in both transverse planes. c) images representing the evolution of the 2D transverse 1+ beam profile along the SP1CB.

*Blind tuning test*

As it is important to note, RIB facilities need to deal with beams which are orders of magnitude lower in terms of intensity than stable beams. Classical faraday cups cannot be used to tune the beam optic elements when dealing with such intensities. The typical method to tune the radioactive beams, is to use a stable one with an intensity in the range of hundred nA or more. Once tuned on the stable beam, one parameters of the beam line one are scaled for transporting the RIB, with respect to Bρ for the magnetic optical elements, and with respect to 1+ platform potential for the electrostatic optical elements. for.

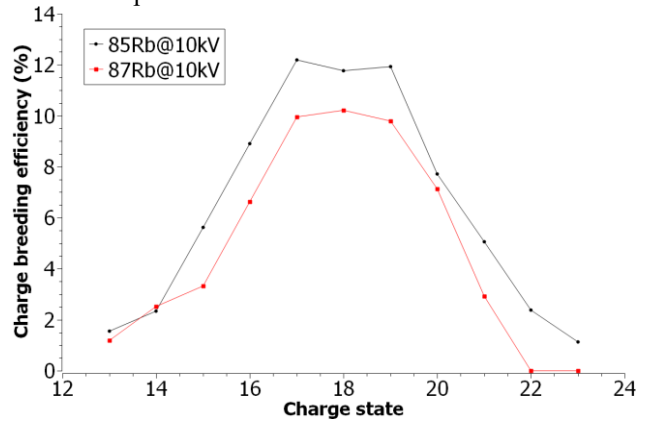

Figure 11: Charge state distribution of $^{85}$Rb (black) and $^{87}$Rb (red) charge bred ions at 10keV of total 1+ beam energy.

Such test has been accomplished successfully using the two stable isotopes of rubidium: $^{85}$Rb (stable) and $^{87}$Rb (stable like because its half-life is 4,92×10$^{10}$ years). The abundances of $^{85}$Rb and $^{87}$Rb are ~72.2% and 27.8% respectively. The ion gun delivered ion beam intensities of ~300 nA of $^{85}$Rb and 120 nA of $^{87}$Rb well mass resolved. The test has been done at acceleration voltage of 10 kV and using He as buffer gas for the SP1CB. The full SPIRAL1 was tuned for getting a charge bred beam of $^{85}$Rb$^{19+}$ (Brho = 0.0305 T.m) from $^{85}$Rb$^{1+}$ (Brho = 0.1332 T.m). At the same acceleration voltage, the Bρ's of the $^{87}$Rb$^{1+}$ and $^{87}$Rb$^{19+}$ ion beam become 0.1347 T.m and 0.0309 T.m (ratio of square root of the masses $\sqrt{\frac{87}{85}}$) meaning a factor of 1.0117. Figure 11 shows the CSD of charge bred ions. They are both peaked on the $^{85}$Rb$^{17+, 18+, 19+}$ ions with a maximum charge breeding efficiency of 12.2% ($^{85}$Rb$^{17+}$) and 10.2% ($^{87}$Rb$^{18+}$) associated with global efficiency of 70.6% and 53.6% for ($^{85}$Rb) and $^{87}$Rb respectively. The SP1CB as well as 1+ and N+ beam lines were tuned exactly in the same manner. Only the magnetic optical elements were scaled regarding the Bρ's ratio given above. The ΔV value remained constant to 5.4V (ΔV is the additional voltage applied on the ion gun to get enough energy for the Rb$^{1+}$ beam to be efficiently captured by the SP1CB plasma).

That result tends to prove that blind tuning is possible as long as the relative Bρ difference is of the order of few tens of %.

## ROLE OF ENERGY SPREAD

The energy 1+ beam energy acceptance of the SP1CB has been measured to be in of order of few eV, from ~3eV up to ~8eV, [13]. In other terms 1+ ions with an energy ~2eV away with respect to the optimum ΔV will divide by a factor between ~1.3 and 2 their chance to be captured by the core of the ECR plasma. Measurements have been achieved to get the energy spread of the two TISS's presently in use at SPIRAL, and for the ion gun used in the studies presented here. For that purpose, a Retarding Field Analyzer has been used. This device was formerly developed by the LPC Caen laboratory. Its design is very similar to the one presented in ref. [14]. It is composed of an aperture of 1mm of diameter for the incoming beam; an ion collector at the opposite position, two Ni grids to slow down the monocharged ions, all of this being embedded in a metallic cage (see fig.12). The maximal energy of the beam must be 5keV because that design does not accept voltage above 5kV due to the distances and insulators dimensions.

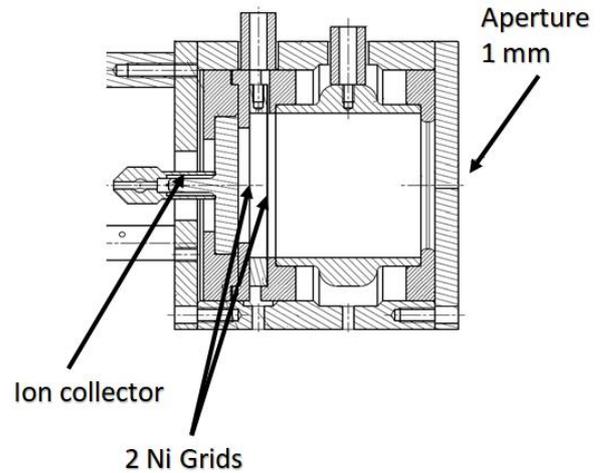

Figure 12: Scheme of the Retarding Field Analyzer.

The 1+ beam was not cut-off by any slit jaws before the measurement. A simple acquisition system based on Labview ramps up the voltages on grids and the ion current measured on the ion collector is recorded (see Figure 13 Signal measurement). That signal is multiplied by a factor (-1) and is then fitted with a sigmoidal equation (Eq.1) to get a smooth curve. That one is derived to obtain a curve fitted with a Normal distribution. From the sigma of the Normal distribution (see Figure 13 Derivative of the signal), the FWHM is deduced giving finally the energy spread (FWHM ≅ 2.355σ).

$$f(x) = (A1 - A2)/(1 + \exp((x - x0)/dx)) + A2 \quad (Eq. 1)$$

Figure 14 displays the results found for the three 1+ ion sources. The ion gun being a thermo-ionization source, its energy spread is rather low: 2.17±0.01 eV. The FEBIAD and NanoganIII sources are based on plasma with higher energy spread of 3.93±0.02eV and 4.79±0.03eV respectively. The latter values agree reasonably well with previously reported energy profiles for these sources [9]. The charge breeding efficiencies measured so far with the different ion sources show a large disparity: they are within the range of 10-16% (e.g. $^{85}$Rb$^{18+}$ & $^{23}$Na$^{7+}$) for the Ion gun; of ~2 - 5% ($^{38m}$K$^{8+}$ and $^{37}$K$^{9+}$) for the TISS FEBIAD, for which only a limited time/experience for the 1+ to N+ beam tuning has been available; and of ~1% ($^{32}$S$^{7+}$ & $^{19}$F$^{5+}$) for the NanoganIII TISS.

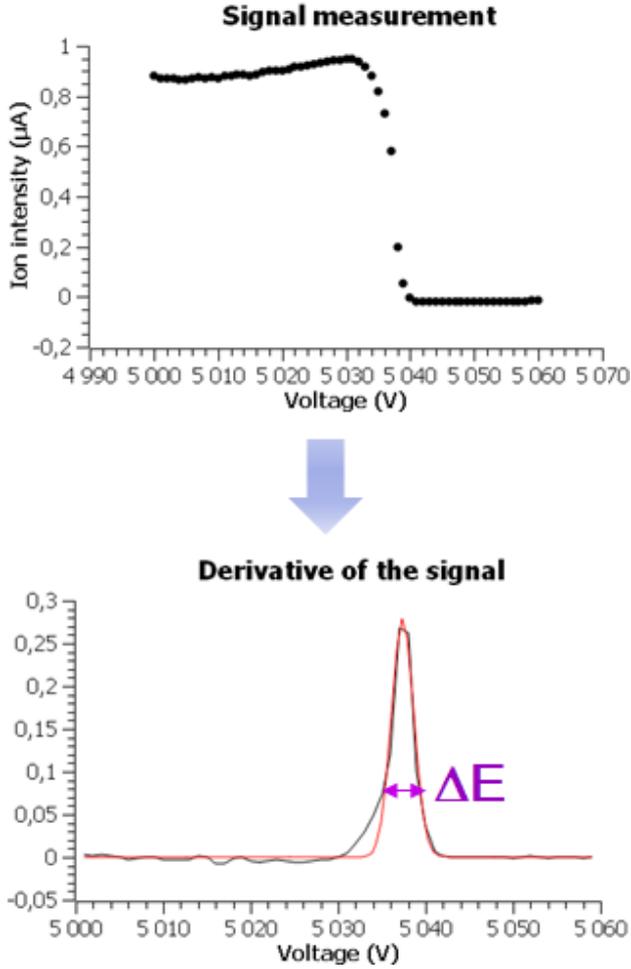

Figure 13: Scheme of the Retarding Field Analyser.

One may therefore wonder what are the main responsible factors in this disparity. Among the possible factors, the most probable ones are differences in i) 1+ beam energy spread, ii) 1+ beam emittance iii) conditions of the 1+ beam optics tuning. Table 1 permits a comparison of these different factors for the diverse sources. The proportion of the accepted 1+ beam with regards of the energy acceptance of the booster is estimated in the 4th column from the left, as a simple convolution product of two Gaussian functions. As it appears from the table, the difference in energy spread is unlikely to explain alone the variation in charge breeding efficiencies. The emittances of the ion gun and FEBIAD are similar owing to a very compact geometry and tiny extraction, so that the modest performances with the FEBIAD can only be attributed to the limited time that has been available so-far in the SPIRAL 1 beam lines for tuning the 1+ beam into the charge breeder. Tests are presently pursued with this ion source, which corroborate this interpretation. Efficiencies much more comparable to the ones of the ion gun are being demonstrated for stable metallic elements such as Mg and Fe. The final results of these tests will be presented after a full analysis in a forthcoming publication. The role of the emittance has been demonstrated in ref. [2], and is advocated as the main reason for the modest charge breeding efficiency obtained with the 1+ beams from the NanoganIII source. At this stage, a dedicated study would be necessary to fully quantify the role of the emittance in the charge breeding efficiency. This could be done by eg. controlling the emittance of the 1+ beam from NanoganIII thanks to a system of slits, which is a feature available in the SPIRAL1 1+ beam line.

Table 1: Comparison of measured charge breeding efficiencies for the different sources, and of the possible reasons for their differences. The 4th column from the left corresponds to the calculated proportion of accepted 1+ beam according to an energy acceptance of the booster varying from 3 to 8 eV FWHM.

| Ion source | charge breeding efficiencies (min-max) | Energy spread / eV | Accepted in booster (min - max) | Emittance $2\sigma$ RMS / $\pi$.mm.mrad | Conditions 1+ beam tuning |
|---|---|---|---|---|---|
| Ion gun | 10-16% | 2,17 | 81% - 96% | ~30 | Thoroughly tested |
| FEBIAD | 2-5% | 3,93 | 60% - 89% | ~30 | A few hours |
| NanoganIII | 1% | 4,79 | 52% - 85% | ~100 | A few days |

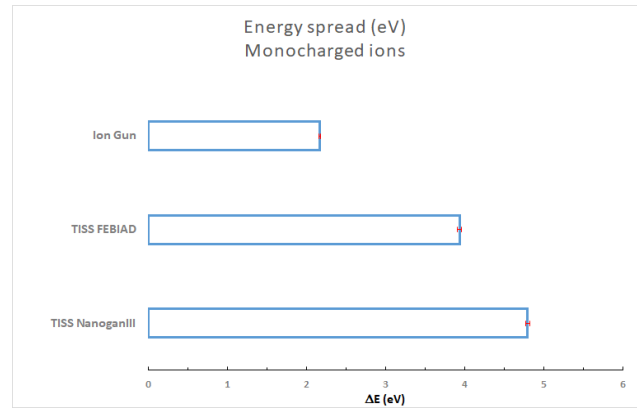

Figure 14: Energy spread measurements for the Ion gun ($^{23}$Na$^+$) and the two TISS's FEBIAD ($^{39}$K$^+$) and NanoganIII ($^{40}$Ar$^+$)

## CONCLUSION

That paper focuses on the beam optics of the low energy beams of the SPIRAL 1 facility. Concerning the ST mode, more work and diagnostic is requested to better understand why the magnetic field of the SP1CB needs to be decreased in TraceWin by a factor 0.6 to reproduce the experimental beam profiler values. A pepper-pot emittance meter type will be mounted in the N+ beam line to constrain the TraceWin simulation. One additional work to be achieved is to search for a possible optimization that would correlate the settings of the three electrostatic quadrupoles and with the magnetic field created by the SP1CB coils such that the transmission across the SP1CB plasma electrode can be increased.

About the 1+/N+ mode, a new beam optics has been found to overcome the limitation of the magnetic quadrupoles (Q11 Q12 Q13) designed previously to go up to $B\rho$ of 0.136 T.m instead of the current maximum $B\rho$ of 0.22

T.m. That new beam optics leads to performances of the SP1CB close the original one, showing a similar propagation of the ion beam inside the SP1CB. It has been established that both the deceleration electric field and the magnetic field of the booster are required to transport 1+ incoming ions to the core of the SP1CB plasma. Next work will be the validation of that new beam optics with several ion beams including Na and K at few different energies, and covering the full range of Bρ's available at the SPIRAL1 facility. Finally, the final validation will be achieved using the TISS FEBIAD in the 1+/N+ mode. As the FEBIAD source emittance and energy spread are comparable to the ones of the ion gun, similar performances are expected.

## ACKNOWLEDGEMENTS

The authors would like to thanks the fruitful discussions with D$^r$ Olli Tarvainen (STFC Rutherford Appleton Laboratory, ISIS, UK) leading to new insights reported in this paper. Also, the authors want to thank Mr Clément Michel and Mr Alexandre Gognat (GANIL, France) for their computer-assisted design works beneficial to the simulation achievements presented in this paper.